\documentclass[eqsecnum,showpacs,pre,twocolumn]{revtex4}

\usepackage{graphicx}
\usepackage{epsfig}
\usepackage{dcolumn}
\usepackage{amsfonts}
\usepackage{amsmath}
\usepackage{amssymb}
\usepackage{bm}

\begin{document}

\newcommand{\brm}[1]{\bm{{\rm #1}}}
\newcommand{\rsc}{$\mbox{RRN}^{\text{SC}} \hspace{1mm}$}
\newcommand{\rscP}{$\mbox{RRN}^{\text{SC}}$}

\title{Conductivity of continuum percolating systems}

\author{Olaf Stenull}
\author{Hans-Karl Janssen}
\affiliation{
Institut f\"{u}r Theoretische Physik 
III\\Heinrich-Heine-Universit\"{a}t\\Universit\"{a}tsstra{\ss}e 1\\
40225 D\"{u}sseldorf\\
Germany
}

\date{\today} 

\begin{abstract}
We study the conductivity of a class of disordered continuum systems represented by the Swiss-cheese model, where the conducting medium is the space between randomly placed spherical holes, near the percolation threshold. This model can be mapped onto a bond percolation model where the conductance $\sigma$ of randomly occupied bonds is drawn from a probability distribution of the form $\sigma^{-a}$. Employing the methods of renormalized field theory we show to arbitrary order in $\varepsilon$-expansion that the critical conductivity exponent of the Swiss-cheese model is given by $t^{\text{SC}} (a) = (d-2)\nu + \max [\phi, (1-a)^{-1}]$, where $d$ is the spatial dimension and $\nu$ and $\phi$ denote the critical exponents for the percolation correlation length and resistance, respectively. Our result confirms a conjecture which is based on the 'nodes, links, and blobs' picture of percolation clusters.
\end{abstract}
\pacs{64.60.Ak, 05.60.-k, 72.80.Ng}

\maketitle

\section{Introduction}
\label{intro}
Percolation~\cite{bunde_havlin_91_etc} is one of the best studied problems in statistical physics, both because of its fundamental nature and its vast array of applications. The most natural type of percolation, perhaps, is continuum percolation, where the positions of the constituting elements are not restricted to the discrete sites or bonds of a regular lattice. A simple example of continuum percolation is a conducting material with uniformly-sized holes placed at random. Due to its similarity to Swiss cheese, this model is commonly called the Swiss-cheese model. 

Since the holes are allowed to overlap, the system ceases to support electrical transport when the total volume of the holes exceeds a critical fraction $q_c$. Near this percolation threshold $q_c$ the conducting network consists of many narrow bottlenecks each of which is bounded by inter-penetrating holes. Thus, it is plausible that the Swiss-cheese model can be mapped onto the random resistor network (RRN) problem where conducting nearest neighbor bonds on a hyper-cubic $d$-dimensional lattice are randomly occupied with a probability $p$. Apparently, the bottlenecks are playing a role similar to the occupied conducting bonds. However, the bottlenecks have a wide distribution of widths, in contrast to the standard RRN, where all occupied bonds are identical. Due to the wide distribution of neck-widths the Swiss-cheese model corresponds to a modified RRN in which the conductances $\sigma$ of the individual occupied bonds have a broad distribution in the form of a power law $\sigma^{-a}$ with $0<a<1$~\cite{halperin&co_85_87}. Due to its relation to the Swiss-cheese model, we abbreviate such a RRN by $\mbox{RRN}^{\text{SC}}$.

It is well established that the purely geometrical percolation exponents for the Swiss-cheese model are in conformity with their analogs in the discrete models~\cite{elam&co_84}. For example, the correlation length $\xi$ is governed in both models by the same exponent $\nu$. The reason is that only the connectivity of the bottlenecks is relevant for the geometrical exponents. As the widths of the individual bottlenecks do not matter, they can be regarded identical in the context of connectivity properties and the problem is essentially equivalent to standard discrete percolation.

The situation is different for critical exponents pertaining to transport quantities. Let us consider the conductivity exponent $t$ for the RRN. It describes the decrease of the average macroscopic conductivity $\Sigma$ when the critical occupation probability $p_c$ is approached from above~\cite{last_thouless_71}
\begin{eqnarray} 
\Sigma \sim (p-p_c )^t \, .
\end{eqnarray}
The conductivity exponent is related to the resistance exponent $\phi$ governing the average resistance $M_R$ between two terminal sites $x$ and $x^\prime$ known to be on the same cluster~\cite{harris_fisch_77,dasgupta_harris_lubensky_78}
\begin{eqnarray}
M_R (x,x^\prime ) \sim \left| x - x^\prime \right|^{\phi /\nu}
\end{eqnarray}
via the scaling relation
\begin{eqnarray}
\label{scaleRelT}
t = (d-2)\nu + \phi \, .
\end{eqnarray}
The conductivity exponent for the \rsc on the other hand depends on a, i.e.,
\begin{eqnarray} 
\Sigma^{\text{SC}} \sim (p-p_c )^{t^{\text{SC}} (a)} \, .
\end{eqnarray}
Early estimates of $t^{\text{SC}} (a)$ were given by Kogut and Straley~\cite{kogut_straley_79}, and Ben-Mizrahi and Bergman~\cite{ben-mizrahi_bergman_81}. Later Straley~\cite{straley_82} argued based on the 'nodes, links, and blobs' picture~\cite{nlb} of percolation clusters that $t^{\text{SC}} (a)$ is given by
\begin{eqnarray}
\label{tVonA}
t^{\text{SC}} (a) = (d-2)\nu + \max \left[ \phi , (1-a)^{-1} \right] 
\end{eqnarray}
(see also Machta {\em et al.}~\cite{machta&co_86}). Without relying on the assumptions of the 'nodes, links, and blobs' picture $t^{\text{SC}} (a)$ has been addressed by Lubensky and Tremblay (LT)~\cite{lubensky_tremblay_86} from a renormalization group (RG) perspective. After some controversy~\cite{machta_88&lubensky_tremblay_88} their perturbation calculation to first order in the deviation from the upper critical dimension for percolation $\varepsilon = 6-d$ shows agreement with Eq.~(\ref{tVonA}).

The paper in hand presents our field theoretic study of the conductivity of the \rsc. Our analysis builds up on the field theoretic \rsc Hamiltonian by LT. We discuss the RG flow for the whole regime $0 < a <1$ to arbitrary order in a diagrammatic expansion. The central result of our work is that Eq.~(\ref{tVonA}) holds to arbitrary order in $\varepsilon$. 

The outline of the remainder of this paper is the following. Section~\ref{theModel} describes the modeling. We define the percolation problem under consideration. Then we show how the average resistance and the related average conductance can be derived from a generating function. We explain how the replica trick facilitates averaging and leads to an effective Hamiltonian. Next, this effective Hamiltonian is refined into a field theoretic functional. A scaling analysis concludes Sec.~\ref{theModel}. It reveals the relevance of the field theoretic couplings associated with the conductances of the occupied bonds. In Sections~\ref{RGA_RRN} and~\ref{RGA_RSC} we actually compute the generating function for the average conductance by employing field theory augmented by renormalization. A Gell--Mann-Low RG equation (RGE) provides us with the scaling behavior of the average conductance near criticality. Our analysis is partitioned into two cases. In Sec.~\ref{RGA_RRN} we consider $a=0$ and basically review the known results for the RRN. Section~\ref{RGA_RSC} deals with the case of prime interest, viz.\ $0<a<1$.  Finally, concluding remarks are given in Section~\ref{concusions}. Technical details are relegated to Appendices~\ref{app:XXX} and \ref{app:YYY}.

\section{The model}
\label{theModel}
We are about to consider a bond percolation model on a $d$-dimensional hyper-cubic lattice where the conductances of the occupied bonds are independently and identically distributed random variables. To be specific, the distribution function $\mathfrak{g}$ of the conductance $\sigma_b$ of any bond $b$ is taken to be
\begin{subequations}
\begin{eqnarray}
\label{gDef}
\mathfrak{g} \left( \sigma \right) = \left( 1-p \right) \delta \left( \sigma \right) + p \, \mathfrak{f} \left( \sigma \right)
\end{eqnarray}
where
\begin{eqnarray}
\label{fDef}
\mathfrak{f} \left( \sigma \right) = \left( 1-a \right) \sigma_0^{-1} \left( \frac{\sigma}{\sigma_0} \right)^{-a}
\end{eqnarray}
\end{subequations}
with $\sigma \in [0,\sigma_0]$ and $0<a<1$. For the relation of the \rsc defined by this choice to continuum percolation we refer to Halperin {\em et al}.~\cite{halperin&co_85_87}. Note that $\mathfrak{f}$ has the important feature that the average resistance of an occupied bond is infinite~\cite{footnote1}. This is a key distinction to the standard RRN and also to a RRN with noise modeled by a narrow distribution of bond conductances (cf., e.g., Ref.~\cite{stenull_janssen_2000a&stenull_janssen_2001} and references therein). 

Since we are going to calculate the conductivity exponent $t^{\text{SC}} (a)$ via the average conductance $M_{R^{-1}}^{\text{SC}}$ we need a precise definition of this quantity. Commonly this definition is based on a setup in which a fixed external current $I$ is applied between two leads at lattice sites $x$ and $x^\prime$ known to be on the same cluster. In this setup one could measure the resistance $R(x, x^\prime )$ between the two terminals and then average with respect to $\mathfrak{g}$,
\begin{eqnarray}
\label{defAveConductance}
M_{R}^{\text{SC}} (x, x^\prime )= \left\langle R(x, x^\prime ) \right\rangle_\mathfrak{g}^\prime 
\end{eqnarray}
with the average being defined as
\begin{eqnarray}
\left\langle \cdots \right\rangle_\mathfrak{g}^\prime = \frac{\langle \chi(x, x^\prime ) \cdots \rangle_\mathfrak{g}}{\langle \chi(x, x^\prime )\rangle_\mathfrak{g}} \, .
\end{eqnarray}
Here $\chi(x, x^\prime )$ is an indicator function that takes on the value one if $x$ and $x^\prime$ are located on the same cluster and zero otherwise. The average macroscopic resistance (\ref{defAveConductance}), however, has the severe drawback that it is not well defined, because the average bond resistance diverges. Thus, it is preferable to work with the average conductance instead which is given by
\begin{eqnarray}
M_{R^{-1}}^{\text{SC}} (x, x^\prime ) = \left\langle R(x, x^\prime )^{-1} \right\rangle_\mathfrak{g}^\prime \, .
\end{eqnarray}

Note that $\chi$ probes only geometrical connectivity. Hence, $\langle \chi(x, x^\prime )\rangle_\mathfrak{g}$ can be identified with $\langle \chi(x, x^\prime )\rangle_C$, where $\langle \cdots \rangle_C$ denotes averaging over all diluted lattice configurations $C$ of the corresponding standard bond percolation model. Accordingly $\langle \chi(x, x^\prime )\rangle_\mathfrak{g}$ is nothing more than the usual percolation correlation function, i.e., the probability $P (x, x^\prime)$ that $x$ and $x^\prime$ are connected. 

\subsection{Generating function}
\label{replicaFormalism}
In this section we review how one can devise a generating function for $M_{R^{-1}}^{\text{SC}}$ based on the ideas of Stephen~\cite{stephen_78}. We demonstrate that this generating function indeed serves its purpose and explain how the average conductance can be extracted from it.

Stephen introduced the quantity
\begin{eqnarray}
\label{defPsi}
\psi_{\vec{\lambda}}(x) = \exp \left( i \vec{\lambda} \cdot \vec{V}_x \right) \, , \quad \vec{\lambda} \neq \vec{0} \, .
\end{eqnarray}
$\vec{V}_x = ( V_x^{(1)}, \cdots , V_x^{(D)} )$ is a $D$-fold replicated variant of the voltage $V_x$ at lattice site $x$ and $\vec{\lambda} = ( \lambda^{(1)}, \cdots , \lambda^{(D)} )$ is, apart from a factor $-i$, a replicated external current. The corresponding scalar product is defined as $\vec{\lambda} \cdot \vec{V}_x = \sum_{\alpha =1}^D V_x^{(\alpha )} \lambda^{(\alpha )}$. The physical content of $\psi_{\vec{\lambda}}(x)$ will be explained below. 

In order to proceed towards the desired generating function we now consider the two-point correlation function of $\psi_{\vec{\lambda}}(x)$
\begin{eqnarray}
\label{correlPsi}
G \left( x, x^\prime ,\vec{\lambda} \right) = \left\langle 
\psi_{\vec{\lambda}}(x)\psi_{-\vec{\lambda}}(x^\prime) 
\right\rangle_{\mbox{\scriptsize{rep}}}
\end{eqnarray}
where the average is defined by
\begin{eqnarray}
\label{erzeugendeFunktion}
\Big\langle \cdots \Big\rangle_{\mbox{\scriptsize{rep}}} &=& \bigg\langle Z^{-D} \int \prod_j \prod_{\alpha =1}^D dV_j^{(\alpha )} 
\nonumber \\
&\times&
\exp \bigg[ -\frac{1}{2} P \left( \left\{ \vec{V} \right\} \right)  \bigg] \cdots \bigg\rangle_\mathfrak{g} \, .
\end{eqnarray}
The product over $j$ is taken over all lattice sites. $Z$ is a normalization factor given by
\begin{eqnarray}
\label{norm}
Z = \int \prod_{j} dV_{j} \exp \left[ -\frac{1}{2} P \left( \left\{ V \right\} \right) \right] \, .
\end{eqnarray}
$P ( \{ V \} )$ denotes the dissipated power
\begin{eqnarray}
P ( \{ V \} ) = \sum_{b} \sigma_{b} V_b^2 = \sum_{<i,j>} \sigma_{i,j} \left( V_i - V_j \right)^2
\end{eqnarray}
with the summations running over all bonds. $V_b$ abbreviates $V_i - V_j$, where $i$ and $j$ are the lattice sites belonging to the respective bond, and accordingly, $\sigma_{i,j} = \sigma_{b}$. $P ( \{ \vec{V} \} )$ is the replicated version of the power with all voltages replaced by their replicated analogs.

Before evaluating Eq.~(\ref{correlPsi}) we need to comment on regularization issues. First, it is important to realize that the integrands in Eqs.~(\ref{erzeugendeFunktion}) and (\ref{norm}) depend only on voltage differences and hence the integrals are divergent. To give these integrals a well defined meaning one can introduce an additional power term $\frac{i\omega}{2} \sum_i V^2_i$. Physically the new term corresponds to grounding each lattice site by a capacitor of unit capacity. The original situation may be restored by taking the limit of vanishing frequency, $\omega \to 0$. Second, it is not guaranteed that $Z$ stays finite because infinite voltage drops may occur. Thus, the limit $\lim_{D \to 0}{Z^D}$ is not well defined. This problem can be regularized by switching to voltage variables $\vec{\theta}$ taking discrete values  on a $D$-dimensional torus that we refer to as the replica space. The voltages are discretized by setting $\vec{\theta} = \Delta \theta \vec{k}$, where $\Delta \theta = \theta_M /M$ is the gap between successive voltages, $\theta_M$ is a voltage cutoff, $\vec{k}$ is a $D$-dimensional integer, and $M$ a positive integer. The components of $\vec{k}$ are restricted to $-M < k^{(\alpha)} \leq M$ and periodic boundary conditions are realized by equating $k^{(\alpha )}=k^{(\alpha )} \mbox{mod} (2M)$. The continuum may be restored by taking $\theta_M \to \infty$ and $\Delta \theta \to 0$. By setting $M=m^2$, $\theta_M = \theta_0 \, m$, and, respectively, $\Delta \theta = \theta_0 /m$, the two limits can be taken simultaneously via $m \to \infty$. Note that the limit $D \to 0$ has to be taken before $m \to \infty$ in order to ensure $\lim_{D\to 0} (2M)^{-D} =1$. Since the voltages and $\vec{\lambda}$ are conjugated variables, $\vec{\lambda}$ is affected by the discretization as well:
\begin{eqnarray}
\vec{\lambda} = \Delta \lambda \, \vec{l} \ , \ \Delta \lambda \, \Delta \theta = \pi /M \, ,
\end{eqnarray}
where $\vec{l}$ is a $D$-dimensional integer taking the same values as $\vec{k}$. This choice guarantees that the completeness and orthogonality relations
\begin{subequations}
\label{complete}
\begin{eqnarray}
\frac{1}{(2M)^D} \sum_{\vec{\theta}} \exp \left( i \vec{\lambda} \cdot \vec{\theta} \right) = \delta_{\vec{\lambda} ,\vec{0} 
\hspace{0.15em}\mbox{\scriptsize{mod}}(2M \Delta \lambda) }
\end{eqnarray}
and
\begin{eqnarray}
\frac{1}{(2M)^D} \sum_{\vec{\lambda}} \exp \left( i \vec{\lambda} \cdot \vec{\theta} \right) = \delta_{\vec{\theta} ,\vec{0} 
\hspace{0.15em}\mbox{\scriptsize{mod}}(2M \Delta \theta)}
\end{eqnarray}
\end{subequations}
do hold. Equation~(\ref{complete}) provides us with a Fourier transform between the $\vec{\theta}$- and $\vec{\lambda}$-tori. It is important to note that the replica space Fourier transform of $\psi_{\vec{\lambda}}(x)$, 
\begin{eqnarray}
\label{fouriertransform}
\Phi_{\vec{\theta}} \left( x \right) &=& (2M)^{-D} \sum_{\vec{\lambda} \neq \vec{0}} \exp \left( i \vec{\lambda} \cdot \vec{\theta} 
\right) \psi_{\vec{\lambda}} (x) 
\nonumber \\
&=& \delta_{\vec{\theta}, \vec{\theta}_{x}} - (2M)^{-D} 
\end{eqnarray}
satisfies the condition $\sum_{\vec{\theta }} \Phi_{\vec{\theta}} ( x ) = 0$ and hence is nothing more than a Potts spin~\cite{Zia_Wallace_75} with $q=(2M)^D$ states.

In passing we emphasize the benefit of the replication procedure. It provides us with an extra parameter $D$ that we may tune to zero. In this replica limit the normalization denominator $Z^{-D}$ goes to one and hence does not depend on the distribution of the bond conductances anymore. Then the only remaining dependence on this distribution rests in the power $P$ appearing in the exponential in Eq.~(\ref{erzeugendeFunktion}). In the replica limit, therefore, we just have to average this exponential instead of the entire right hand side of Eq.~(\ref{erzeugendeFunktion}). This average then provides us with an effective power or Hamiltonian which serves as vantage point for all further calculations. The effective Hamiltonian will be discussed in Sec.~\ref{fieldTheoreticHamiltonian}.

Now we come back to the role of Eq.~(\ref{correlPsi}) as a generating function. Since the integrations are Gaussian they are readily carried out with the result
\begin{eqnarray}
\label{GenFktLambda}
G \left( x, x^\prime ,\vec{\lambda} \right) = P \left( x, x^\prime \right) \bigg\langle \exp \bigg[ - \frac{\vec{\lambda}^2}{2} R \left( x,x^\prime \right)  \bigg] \bigg\rangle_\mathfrak{g}^\prime \, .
\nonumber \\
\end{eqnarray}
It is evident from Eq.~(\ref{GenFktLambda}) that $G( x, x^\prime ,\vec{\lambda} )$ represents a generating function for the average resistance. To obtain a generating function for the average conductance one simply needs to carry out a Fourier transformation in replica space,
\begin{eqnarray}
\label{GenFktTheta1}
\widetilde{G} \left( x, x^\prime ,\vec{\theta} \right) &=& P \left( x, x^\prime \right) \frac{1}{(2M)^D} \sum_{\vec{\lambda}} \exp \left( i \vec{\lambda} \cdot \vec{\theta} \right)
\nonumber \\
&\times&
\bigg\langle \exp \bigg[ - \frac{\vec{\lambda}^2}{2} R \left( x,x^\prime \right)  \bigg] \bigg\rangle_\mathfrak{g}^\prime \, .
\end{eqnarray}
After paying due attention to the exclusion of $\vec{\lambda} = \vec{0}$ we may approximate the summation in Eq.~(\ref{GenFktTheta1}) by an integration, 
\begin{eqnarray}
\label{GenFktTheta2}
&&\widetilde{G} \left( x, x^\prime ,\vec{\theta} \right) = P \left( x, x^\prime \right) \frac{1}{(2M\, \Delta \lambda)^D}   
\nonumber \\
&&\times \, \bigg\{
\bigg\langle \int_{-\infty}^\infty d^D \lambda \, \exp \bigg[ - \frac{\vec{\lambda}^2}{2} R \left( x,x^\prime \right) + i \vec{\lambda} \cdot \vec{\theta} \bigg] \bigg\rangle_\mathfrak{g}^\prime\nonumber \\ 
& & - \, \frac{1}{(2M)^D} \bigg\} \, .
\end{eqnarray}
The $\lambda$-integration is straightforward since it is Gaussian. In the limit $D\to 0$ we obtain
\begin{eqnarray}
\label{GenFktTheta3}
&&\widetilde{G} \left( x, x^\prime ,\vec{\theta} \right) = P \left( x, x^\prime \right)   
\nonumber \\ 
&& \times \, \bigg\{
\bigg\langle  \exp \bigg[ - \frac{\vec{\theta}^2}{2} R \left( x,x^\prime \right)^{-1}  \bigg] \bigg\rangle_\mathfrak{g}^\prime\nonumber  - 1 \bigg\} 
\nonumber \\
&& = \, P \left( x, x^\prime \right) \bigg\{ - \frac{\vec{\theta}^2}{2} M_{R^{-1}}^{\text{SC}} \left( x,x^\prime \right) + \cdots \bigg\}
\, .
\nonumber \\
\end{eqnarray}
We learn from Eq.~(\ref{GenFktTheta3}) that $\widetilde{G} ( x, x^\prime ,\vec{\theta} )$ is indeed the generating function we are looking for and that $ M_{R^{-1}}^{\text{SC}}$ can be extracted simply by taking the appropriate derivative,
\begin{eqnarray}
\label{melkeGenFkt}
M_{R^{-1}}^{\text{SC}} \left( x,x^\prime \right) = P \left( x, x^\prime \right)^{-1} \frac{\partial}{\partial ( - \vec{\theta}^2/2 )} \widetilde{G} \left( x, x^\prime ,\vec{\theta} \right) \Big|_{\vec{\theta}^2 =0}
\, .
\nonumber \\
\end{eqnarray}

We conclude Sec.~\ref{replicaFormalism} by addressing the physical content of $\psi_{\vec{\lambda}}(x)$ and its replica space Fourier transform $\Phi_{\vec{\theta}}(x)$. A reasoning similar to that for the two-point correlation function $G ( x, x^\prime ,\vec{\lambda} )$ leads to
\begin{eqnarray}
\label{psiAve}
\left\langle \psi_{\vec{\lambda}}(x) \right\rangle_{\mbox{\scriptsize{rep}}} = P_\infty \bigg\langle \exp \bigg[ - \frac{\vec{\lambda}^2}{4} R_\infty \left( x \right)  \bigg] \bigg\rangle_\mathfrak{g}^\prime \, ,
\end{eqnarray}
where the prime now indicates averaging subject to the condition that $x$ is located on an infinite cluster. $P_\infty$ stands for the percolation probability that a point belongs to an infinite cluster and $R_\infty (x)$ denotes the resistance between $x$ and infinity. From Eq.~(\ref{psiAve}) we learn an important feature of $\psi_{\vec{\lambda}}(x)$, namely that its average is proportional to the percolation order parameter $P_\infty$. For reasons that are clear by now it is preferable to consider $\Phi_{\vec{\theta}}(x)$. Upon Fourier transformation in replica space we find in the limit $D\to 0$
\begin{eqnarray}
\label{phiAve1}
\left\langle \Phi_{\vec{\theta}}(x) \right\rangle_{\mbox{\scriptsize{rep}}} = P_\infty \bigg\{ \bigg\langle \exp \bigg[ - \vec{\theta}^2 R_\infty \left( x \right)^{-1}  \bigg] \bigg\rangle_\mathfrak{g}^\prime 
- 1 \bigg\} \, .
\nonumber \\
\end{eqnarray}
Anticipating results we will derive in Sec.~\ref{RGA_RSC} we rewrite Eq.~(\ref{phiAve1}) as
\begin{eqnarray}
\label{phiAve2}
\left\langle \Phi_{\vec{\theta}}(x) \right\rangle_{\mbox{\scriptsize{rep}}} &=& P_\infty \bigg\{ \int_0^\infty dt \, \mathfrak{p} (t) 
\nonumber \\
&& \times \,
\exp \bigg[ -t  \frac{\vec{\theta}^2}{w \xi^{\phi^{\text{SC}}(a) /\nu}}   \bigg] 
- 1 \bigg\} \, ,
\nonumber \\
\end{eqnarray}
where $w$ is a constant proportional to $\sigma_0^{-1}$ and where
\begin{eqnarray}
\mathfrak{p}(t) = \Big\langle \delta \left( t - w \xi^{\phi^{\text{SC}}(a) /\nu} R_\infty^{-1}\right)  \Big\rangle_\mathfrak{g}^\prime 
\end{eqnarray}
is the probability distribution of the conductance to infinity. Thus, the physical meaning of the averaged $\Phi_{\vec{\theta}}(x)$ may be stated as follows: $\langle \Phi_{\vec{\theta}}(x) \rangle_{\text{rep}}$ corresponds to the percolation order parameter times a scaling function that incorporates the distribution of the conductance to infinity.

\subsection{Field theoretic Hamiltonian}
\label{fieldTheoreticHamiltonian}
This section presents our derivation of a field theoretic Hamiltonian for the \rscP. It is guided be the work of LT. 

We start by revisiting Eq.~(\ref{erzeugendeFunktion}) from which we read off the effective Hamiltonian announced in Sec.~\ref{replicaFormalism}:
\begin{eqnarray}
\label{effHamil1}
H_{\text{rep}} &=&  - \ln \left\langle  \exp \left[ - \frac{1}{2} P \left( \left\{ \vec{\theta} \right\} \right) \right] \right\rangle_\mathfrak{g} 
\nonumber \\
&=& - \ln \bigg\{ \int_0^{\sigma_0} \prod_{b} d \sigma_b \, \mathfrak{g} \left( \sigma_{b} \right)  
\exp \left[ - \frac{1}{2} P \left( \left\{ \vec{\theta} \right\} \right) \right] \bigg\} \, .
\nonumber \\
\end{eqnarray}
For the subsequent steps it is convenient to recast Eq.~(\ref{effHamil1}) as
\begin{eqnarray}
\label{effHamil2}
H_{\mbox{\scriptsize{rep}}} = \sum_{b} K \left( \vec{\theta}_{b} \right)  \, .
\end{eqnarray}
Here, we have introduced
\begin{eqnarray}
\label{kern1}
K \left( \vec{\theta}\right) = - \ln \bigg\{ 1 + \upsilon \int_0^{\sigma_0}  d \sigma \, \mathfrak{f} \left( \sigma \right) 
\exp \left[ - \frac{1}{2} \sigma \vec{\theta}^2 \right] \bigg\} \, .
\nonumber \\
\end{eqnarray}
with $\upsilon = p/(1-p)$. Moreover, we dropped a constant $N_B \ln (1-p)$ with $N_B$ being the number of bonds in the undiluted lattice. In order to refine $H_{\text{rep}}$ towards a field theoretic Hamiltonian we now expand $K (\vec{\theta})$ in terms of $\psi_{\vec{\lambda}}(x)$:
\begin{eqnarray}
\label{kern2}
K \left( \vec{\theta}_b \right) 
&=& \frac{1}{\left( 2M \right)^D} \sum_{\vec{\lambda}} \sum_{\vec{\theta}} \exp \left[ i \vec{\lambda} \cdot \left( \vec{\theta}_b - \vec{\theta} \right) \right] K \left( \vec{\theta} \right)
\nonumber \\
&=& \sum_{\vec{\lambda} \neq \vec{0}} \psi_{\vec{\lambda}} \left( i \right) \psi_{-\vec{\lambda}} \left( j \right) 
\widetilde{K} \left( \vec{\lambda} \right) \, ,
\end{eqnarray}
where $\widetilde{K} ( \vec{\lambda} )$ is the replica space Fourier transform of $K (\vec{\theta})$. To evaluate $\widetilde{K} ( \vec{\lambda} )$ we approximate the summation over $\vec{\theta}$ by an integration. This gives, up to a multiplicative factor that goes to one for $D\to 0$, 
\begin{eqnarray}
\widetilde{K} \left( \vec{\lambda} \right) &=& - \int_{-\infty}^{\infty} d^D \theta \exp \left[ i \vec{\lambda} \cdot \vec{\theta} \right] \ln \bigg\{ 1
\nonumber \\
&+& \upsilon \int_0^{\sigma_0}  d \sigma \, \mathfrak{f} \left( \sigma \right) \exp \left[ - \frac{1}{2} \sigma \vec{\theta}^2 \right]
 \bigg\} \, .
\nonumber \\
\end{eqnarray}
Upon expanding the logarithm and carrying out the $\vec{\theta}$-integration we obtain, once more by dropping a multiplicative factor that goes to one in the replica limit,
\begin{eqnarray}
\label{expansionK}
\widetilde{K} \left( \vec{\lambda} \right) = \sum_{l=1}^\infty \frac{(-1)^l}{l}\upsilon^l F_l \left( \vec{\lambda} \right) 
\end{eqnarray}
with $F_l ( \vec{\lambda} )$ being given by
\begin{eqnarray}
\label{vieleInts}
F_l \left( \vec{\lambda} \right) &=& \int_0^{\sigma_0}  d \sigma_1 \cdots \int_0^{\sigma_0}  d \sigma_l \, \mathfrak{f} \left( \sigma_1 \right) \cdots \mathfrak{f} \left( \sigma_l \right) 
\nonumber \\
&\times&
\exp \left[ - \frac{\vec{\lambda}^2}{2} \frac{1}{\sigma_1 + \cdots + \sigma_l} \right] \, .
\end{eqnarray}
Integration yields, as demonstrated in Appendix~\ref{app:XXX},
\begin{eqnarray}
\label{resF}
F_l \left( \vec{\lambda} \right) = 1 + A_l \, \vec{\lambda}^2 + B_l \, \vec{\lambda}^{2l(1-a)} + O \Big( \big( \vec{\lambda}^2 \big)^2 \Big) \, ,
\nonumber \\
\end{eqnarray}
where $\vec{\lambda}^{2l(1-a)}$ is understood as $( \vec{\lambda}^2 )^{l(1-a)}$. $A_l$ and $B_l$ are constants. For example, $A_1$ is given by $A_1 = (1-a)/(2 a \sigma_0 )$ and $B_1$ reads $B_1 = - \Gamma (a) /(2\sigma_0 )^{1-a}$ with $\Gamma$ denoting the $\Gamma$-function. The general form of the $A_l$ is $A_l \sim \sigma_0^{-1}[1 + l(a-1)]^{-1}$. For reasons that will be given in Sec.~\ref{scalingAnalysisVoltageVariables} we will neglect all terms associated with $B_{l>1}$ from now on. By inserting Eq.~(\ref{resF}) into Eq.~(\ref{expansionK}) we find
\begin{eqnarray}
\label{resK}
\widetilde{K} \left( \vec{\lambda} \right) = K + w \vec{\lambda}^2 + v \vec{\lambda}^{2l(1-a)} + \cdots 
\end{eqnarray}
with $K$, $w$, and $v$ being expansion coefficients. $v$ is proportional to $\sigma_0^{a-1}$ and positive for $a>0$. $w$ is proportional to $\sigma_0^{-1}$. Its sign depends on the values of $a$ and $\upsilon$. From now on we omit factors $(2M)^{-D}$ that go to one in the replica limit. Moreover, we define the discrete gradient $\nabla_{\vec{\theta}}$ via
\begin{eqnarray}
 - \sum_{\vec{\theta}} \Phi_{\vec{\theta}} \left( i \right) \nabla_{\vec{\theta}}^2 \, \Phi_{\vec{\theta}} \left( j \right) = \sum_{\vec{\lambda} \neq \vec{0}} \vec{\lambda}^2 \psi_{\vec{\lambda}}(i) \psi_{-\vec{\lambda}}(j) \, .
\nonumber \\
\end{eqnarray}
Collecting we find that
\begin{eqnarray}
\label{finalHrep}
H_{\text{rep}} = \sum_{<i,j>} \sum_{\vec{\theta}} \Phi_{\vec{\theta}} \left( i \right) \Big[ K - w  \nabla_{\vec{\theta}}^2  
+ v \left( - \nabla_{\vec{\theta}}^2 \right)^{1-a} \Big] \Phi_{\vec{\theta}} \left( j \right) \, .
\nonumber \\
\end{eqnarray}

In the limit of perfect transport, $\sigma_0 \to \infty$, the coefficients $w$ and $v$ vanish and $H_{\text{rep}}$ reduces to
\begin{eqnarray}
H_{\text{rep}} = K \sum_{<i,j>} \sum_{\vec{\theta}} \Phi_{\vec{\theta}} \left( i \right) \Phi_{\vec{\theta}} \left( j \right) \, .
\end{eqnarray}
This Hamiltonian represents nothing more than the $\left( 2M \right)^D$ states Potts model that is invariant against all $\left( 2M \right)^D !$ permutations of the spin states. If $\sigma_0^{-1} \neq 0$, this $S_{\left( 2M \right)^D}$ symmetry is lost in favor of an $O(D)$ rotational symmetry in replica space. 

We proceed with the usual coarse graining step and replace the Potts spins $\Phi_{\vec{\theta}} ( x )$ by order parameter fields $\varphi ( {\rm{\bf x}} ,\vec{\theta} )$ that inherit the constraint $\sum_{\vec{\theta}} \varphi ( {\rm{\bf x}} ,\vec{\theta} ) = 0$. We model the corresponding field theoretic Hamiltonian $\mathcal{H}$ in the spirit of Landau as a mesoscopic free energy. The constituting elements are local monomials of the order parameter field and its gradients in real and replica space. Purely local terms in replica space have to respect the full $S_{\left( 2M \right)^D}$ Potts symmetry. After these remarks we write down the Landau-Ginzburg-Wilson type Hamiltonian
\begin{subequations}
\label{finalHamil}
\begin{eqnarray}
\mathcal{H} &=& \int d^dx \sum_{\vec{\theta}} \bigg\{ \frac{1}{2} \varphi \left( {\rm{\bf x}} , \vec{\theta} \right) K \left( \Delta ,\Delta_{\vec{\theta}} \right) \varphi \left( {\rm{\bf x}} , \vec{\theta} \right) 
\nonumber \\
&-& \frac{g}{6}\varphi \left( {\rm{\bf x}} , \vec{\theta} \right)^3 \bigg\} \, , \end{eqnarray}
with the kernel being given by
\begin{eqnarray}
 K \left( \Delta ,\Delta_{\vec{\theta}} \right) = \tau - \nabla^2 - w  \nabla_{\vec{\theta}}^2 + v \left( - \nabla_{\vec{\theta}}^2 \right)^{1-a} \, .
\nonumber \\
\end{eqnarray}
\end{subequations}
In Eq.~(\ref{finalHamil}) we have neglected all higher order terms that are irrelevant in the renormalization group sense. $w$, and $v$ are now coarse grained analogues of the original coefficients appearing in Eq.~(\ref{finalHrep}). The parameter $\tau - \tau_c \sim (p_{c}-p)$ specifies the deviation of the occupation probability $p$ from the critical probability $p_{c}$. In mean field theory the percolation transition happens at $\tau = \tau_c =0$. We point out that $\mathcal{H}$ reduces to the usual field theoretic Hamiltonian for the $(2M)^D$ states Potts model upon setting $w=v=0$. Thus, $\mathcal{H}$ satisfies an important consistency requirement since one retrieves purely geometrical percolation in the limit $\sigma_0 \to \infty$.

\subsection{Scaling analysis in the voltage variable}
\label{scalingAnalysisVoltageVariables}
Now we address the relevance of the coupling constants associated with the bond conductances. We carry out a scaling analysis by rescaling the voltage variable: $\vec{\theta} \to b \vec{\theta}$. Here $b$ denotes a scaling factor and should not be confused with the index labeling the bonds. By substituting $\varphi ( {\rm{\bf x}} , \vec{\theta}) = \varphi^\prime ( {\rm{\bf x}} , b\vec{\theta} )$ into the Hamiltonian we get
\begin{eqnarray}
\label{scaling1}
&&\mathcal{H} \left[ \varphi^\prime \left( {\rm{\bf x}} , b \vec{\theta} \right) ; \tau ,w, v \right] 
\nonumber \\
&& = \, \int d^dx \sum_{\vec{\theta}} \bigg\{ \frac{1}{2} 
\varphi^\prime \left( {\rm{\bf x}} , b \vec{\theta} \right) K \left( \Delta 
,\Delta_{\vec{\theta}} \right) \varphi^\prime \left( {\rm{\bf x}} , b \vec{\theta} \right) 
\nonumber \\
&&
- \, \frac{g}{6}\varphi^{\prime } \left( {\rm{\bf x}} , b \vec{\theta} 
\right)^3 \bigg\} \, .
\end{eqnarray}
Renaming the scaled voltage variables $\vec{\theta}^\prime = b \vec{\theta}$ leads to
\begin{eqnarray}
\label{scaling2}
&&\mathcal{H} \left[ \varphi^\prime \left( {\rm{\bf x}} , \vec{\theta}^\prime \right) ; \tau ,w, v \right] 
\nonumber \\
&& = \, \int d^dx \sum_{\vec{\theta}^\prime} \bigg\{ \frac{1}{2} 
\varphi^\prime \left( {\rm{\bf x}} , \vec{\theta}^\prime \right) K \left( \Delta , b^2\Delta_{\vec{\theta}} \right) \varphi^\prime \left( {\rm{\bf x}} ,\vec{\theta}^\prime \right) 
\nonumber \\
&&
- \, \frac{g}{6}\varphi^{\prime } \left( {\rm{\bf x}} , \vec{\theta}^\prime \right)^3 \bigg\} \, .
\end{eqnarray}
Obviously, a scaling of the voltage variable results in a scaling of the voltage cutoff, $\theta_0 \to b \theta_0$. However, by taking the limit $D \to 0$ and then $m \to \infty$, the dependence of the theory on the cutoff drops out. In other words: $\theta_0$ is a redundant scaling variable. Thus, one can identify $\vec{\theta}^\prime$ and $\vec{\theta}$ and conclude that 
\begin{eqnarray}
\label{relForH}
&&\mathcal{H} \left[ \varphi \left( {\rm{\bf x}} , b \vec{\theta} \right) ; \tau ,w, v \right] 
\nonumber \\
&&=\, \mathcal{H} \left[ \varphi \left( {\rm{\bf x}} , \vec{\theta} \right) ; \tau , b^2 w, b^{2(1-a)} v \right] \, . 
\end{eqnarray}

Next we consider the consequences of Eq.~(\ref{relForH}) for the correlation functions of the field $\varphi ( {\rm{\bf x}} , \vec{\theta} )$ given by
\begin{eqnarray}
\label{correl}
&&\widetilde{G}_N \left( \left\{ {\rm{\bf x}} ,\vec{\theta} \right\} ; \tau , w, v \right) 
\nonumber \\
&& =\, \int \mathcal{D} \varphi \ \varphi \left( {\rm{\bf x}}_1 , \vec{\theta}_1 \right) \cdots \varphi \left( {\rm{\bf x}}_N , \vec{\theta}_N \right)
\nonumber \\
& & \times \exp \left( - \mathcal{H} \left[ \varphi \left( {\rm{\bf x}} , \vec{\theta} \right) ; \tau , w, v \right] \right) \, ,
\end{eqnarray}
where $\mathcal{D} \varphi$ indicates an integration over the set of variables $\{ \varphi ( {\rm{\bf x}} , \vec{\theta} ) \}$ for all ${\rm{\bf x}}$ and $\vec{\theta}$. Equation~(\ref{relForH}) implies that
\begin{eqnarray}
\label{ohnasch}
&&\widetilde{G}_N \left( \left\{ {\rm{\bf x}} ,\vec{\theta} \right\} ; \tau , w, v  \right) 
\nonumber \\
&&=\, \widetilde{G}_N \left( \left\{ {\rm{\bf x}} , b \vec{\theta} \right\} ; \tau , b^2 w, b^{2(1-a)} v \right) \, .
\end{eqnarray}
From Eq.~(\ref{ohnasch}) in conjunction with Eq.~(\ref{GenFktTheta3}) we deduce
\begin{eqnarray}
\label{ohnasch2}
&&\vec{\theta}^2 M^{\text{SC}}_{R^{-1}} \left( \left( {\rm{\bf x}} , {\rm{\bf x}}^\prime \right) ;\tau , w ,v \right) 
\nonumber \\
&&
= \, b^2 \vec{\theta}^2 
M^{\text{SC}}_{R^{-1}} \left( \left( {\rm{\bf x}} , {\rm{\bf x}}^\prime \right) ;\tau , b^2 w , b^{2(1-a)} v \right) \, .
\nonumber \\
\end{eqnarray}

The freedom of choice with respect to $b$ has not been exploited yet. To address the issue of relevance we choose $b^2 = v^{-1/(1-a)}$. This gives
\begin{eqnarray}
\label{ohnasch3}
&& M^{\text{SC}}_{R^{-1}} \left( \left( {\rm{\bf x}} , {\rm{\bf x}}^\prime \right) ;\tau , w ,v \right) 
\nonumber \\
&&
= \, v^{-1/(1-a)}
M^{\text{SC}}_{R^{-1}} \left( \left( {\rm{\bf x}} , {\rm{\bf x}}^\prime \right) ;\tau , w /v^{1/(1-a)} , 1 \right) \, .
\nonumber \\
\end{eqnarray}
By virtue of $v \sim \sigma_0^{-(1-a)}$ we may recast Eq.~(\ref{ohnasch3}) as
\begin{eqnarray}
\label{ohnasch4}
&& M^{\text{SC}}_{R^{-1}} \left( \left( {\rm{\bf x}} , {\rm{\bf x}}^\prime \right) ;\tau , w ,v \right) 
\nonumber \\
&&
= \, \sigma_0^{-1}
f_1 \left( \left( {\rm{\bf x}} , {\rm{\bf x}}^\prime \right) ;\tau , w /v^{1/(1-a)} \right) \, ,
\end{eqnarray}
with $f_1$ being a scaling function. We learn that $w$ appears only in the combination $w /v^{1/(1-a)}$. A trivial consequence of the fact that the Hamiltonian~(\ref{finalHamil}) must be dimensionless is that $w \vec{\lambda}^{2} \sim v \vec{\lambda}^{2(1-a)} \sim \mu^2$, where $\mu$ is an inverse length scale. Thus, $w /v^{1/(1-a)} \sim \mu^{-2a/(1-a)}$. This leads to the conclusion that $w$ is marginal for $a$ of order $\varepsilon$ whereas it is clearly irrelevant for $a$ of order one. 

As hypothesized in Sec.~\ref{fieldTheoreticHamiltonian} the scaling analysis in the voltage variable justifies that we have neglected in the remainder of Eq.~(\ref{resF}) all terms associated with $B_{l>1}$. Suppose that we had retained these terms. Each of them had contributed a term $-v_l ( - \nabla_{\vec{\theta}}^2 )^{l(1-a)}$ to the kernel in Eq.~(\ref{finalHamil}). From the preceding paragraph it is evident, however, that $v_l$ appears in the average conductance only as $v_l / v^l \sim \mu^{2-2l}$. We conclude that keeping the $B_{l>1}$ had produced only irrelevant terms and that neglecting them in studying the leading behavior at the critical point is indeed justified.

For our RG improved perturbation calculation presented in Sec.~\ref{RGA_RSC} it will be helpful to dispose of a coupling that is invariant under $\vec{\theta} \to b \vec{\theta}$. To identify a candidate we revisit Eq.~(\ref{ohnasch2}) and choose $b^2 = w^{-1}$. This leads to
\begin{eqnarray}
\label{ohnasch5}
M^{\text{SC}}_{R^{-1}} \left( \left( {\rm{\bf x}} , {\rm{\bf x}}^\prime \right) ;\tau , w ,v \right) 
= \sigma_0^{-1}
f_2 \left( \left( {\rm{\bf x}} , {\rm{\bf x}}^\prime \right) ;\tau ,h \right) \, ,
\nonumber \\
\end{eqnarray}
with $f_2$ being another scaling function. $h = v/w^{1-a} \sim \mu^{2a}$ turns out to be the sought invariant coupling constant. We will see that it emerges quite naturally in perturbation theory. Hence, we refer to $h$ as effective coupling.

\section{Review of the RRN}
\label{RGA_RRN}
This section here presents a brief review of the model with $a=0$~\cite{harris_lubensky_87a,stenull_janssen_oerding_99,stenull_2000}. We provide the reader with background on the RRN to make the subsequent analysis of the \rsc more digestible.

In this as well as in the following section we utilize ${\mathcal{H}}$ and calculate the generating function $\widetilde{G} ( {\rm{\bf x}}, {\rm{\bf x}}^\prime , \vec{\theta} )$ by employing field theory augmented by renormalization. For background on these methods we refer to Ref.~\cite{amit_zinn-justin}. As soon as we have $\widetilde{G} ( {\rm{\bf x}}, {\rm{\bf x}}^\prime , \vec{\theta} )$ at hand it is a straightforward matter to extract $M_{R^{-1}}^{\text{SC}}$.

For $a=0$ the coupling constants $v$ and $h$ are redundant and can be set to zero. Straightforward dimensional analysis shows that $d_c =6$ is the upper critical dimension and that $\Gamma_2$ and $\Gamma_3$ are the only superficially divergent vertex functions. The diagrammatic elements as constituents of our perturbation calculation are the three-leg vertex $g$ and the principal propagator 
\begin{eqnarray}
G^{\text{bold}} ( \brm{p}, \vec{\lambda} ) = G ( \brm{k}, \vec{\lambda} ) \{ 1 - \delta_{\vec{\lambda}, \vec{0}} \} \, ,
\end{eqnarray}
where $G (\brm{p}, \vec{\lambda}) = (\tau + \brm{p}^2 + w \vec{\lambda}^2)^{-1}$. Due to the factor $\{ 1 - \delta_{\vec{\lambda}, \vec{0}} \}$ that enforces the constraint $\vec{\lambda} \neq \vec{0}$ the principal propagator decomposes in a conducting part $G^{\text{cond}} ( \brm{p}, \vec{\lambda}) = G ( \brm{p}, \vec{\lambda})$ carrying replica currents and an insulating  part $G^{\text{ins}} ( \brm{p}) = G ( \brm{p}, \vec{\lambda})\delta_{\vec{\lambda}, \vec{0}}$ not carrying replica currents. Each principal diagram decomposes into a sum of conducting diagrams consisting of conducting and insulating propagators. 

Our real-world interpretation~\cite{stenull_janssen_oerding_99,janssen_stenull_oerding_99&janssen_stenull_99,stenull_2000,stenull_janssen_2000a&stenull_janssen_2001,stenull_janssen_oerding_2001,janssen_stenull_directedLetter_2000,stenull_janssen_2001_resistance,stenull_janssen_2001_nonlinear,janssen_stenull_2001_vulcanization}, in which the conducting diagrams are viewed as being resistor networks themselves, provides for a powerful and elegant framework to calculate these diagrams. At first we rewrite the propagators in Schwinger parameterization,  
\begin{eqnarray}
G( \brm{p}, \vec{\lambda} ) =  \int_0^\infty ds\, \exp \left[ -s \left( \tau + \brm{p}^2 + w \vec{\lambda}^2 \right)  \right] \, .
\end{eqnarray}
Next we interpret the Schwinger parameters $s$ of the conducting propagators as their resistance. Then we can express the $\vec{\lambda}$-dependent part of any conducting diagram with $P^{\text{cond}}$ conducting propagators in terms of its electric power $P$:
\begin{eqnarray}
\exp \bigg( -w \sum_{i \in P^{\text{cond}}} s_i \vec{\lambda}^2_i \bigg) = \exp \left[ w P \left( \vec{\lambda} , \left\{ \vec{\kappa} \right\} \right) \right] \, .
\nonumber \\
\end{eqnarray}
The summation on the left hand side runs over all conducting propagators. $\vec{\lambda}_i = \vec{\lambda}_i ( \vec{\lambda} , \{ \vec{\kappa} \} )$, where $\vec{\lambda}$ is an external current and $\{ \vec{\kappa} \}$ is a complete set of independent loop currents, denotes the current flowing through conducting propagator $i$. In this representation it is easy to see that the sum over the loop currents is determined by the total resistance $R ( \{ s_i \} )$ of the respective diagram,
\begin{eqnarray}
\label{toEvaluate}
\sum_{\left\{ \vec{\kappa} \right\}} \exp \left[ w P \left( \vec{\lambda} , \left\{ \vec{\kappa} \right\} \right) \right] = \exp \left[ -R \left(  \left\{ s_i \right\} \right) w \vec{\lambda}^2 \right] \, .
\nonumber \\ 
\end{eqnarray}
Carrying out the usual momentum integrations, which is straightforward after completion of squares, and Taylor expansion gives 
\begin{eqnarray}
\label{diagrStructRRN}
I \left( {\rm{\bf p}}^2 , \vec{\lambda}^2 \right) &=& I_P \left( {\rm{\bf p}}^2 \right) - I_W \left( {\rm{\bf p}}^2 \right) w 
\vec{\lambda}^2 + \cdots
\nonumber \\
&=& \int_0^\infty \prod_i ds_i \left[ 1 - R \left(  \left\{ s_i \right\} \right) w \vec{\lambda}^2 + \cdots \right] 
\nonumber \\
&& \times \, 
D \left( {\rm{\bf 
p}}^2, \left\{ s_i \right\} \right) \, ,
\end{eqnarray}
for the overall form of any conducting diagram. $D ( {\rm{\bf p}}^2, \{ s_i \} )$ stands for the usual (Schwinger parameterized) integrand of the corresponding diagram in the standard $\varphi^3$ theory.

The ultraviolet (UV) divergences encountered in computing the diagrams can be handled by resorting to dimensional regularization. In dimensional regularization the UV divergences appear as poles in the deviation $\varepsilon = 6-d$ from $d_c$. These poles may be eliminated from the superficially divergent vertex functions by employing the renormalization scheme
\begin{subequations}
\begin{eqnarray}
\varphi \to \mathaccent"7017{\varphi} &=& Z^{1/2} \varphi \, ,
\\
\tau \to \mathaccent"7017{\tau} &=& Z^{-1} Z_{\tau} \tau  \, ,
\\
w \to \mathaccent"7017{w} &=& Z^{-1} Z_w w \, , 
\\
g \to \mathaccent"7017{g} &=& Z^{-3/2} Z_u^{1/2} G_\varepsilon^{-1/2} u^{1/2} \, \mu^{\varepsilon/2} \, ,
\end{eqnarray}
\end{subequations}
where the $\mathaccent"7017{}$ indicates unrenormalized quantities. The amplitude $G_\varepsilon = (4\pi )^{-d/2}\Gamma (1 + \varepsilon /2)$ is introduced for convenience. $Z$, $Z_\tau$, and $Z_u$ are the usual Potts model $Z$ factors known to three-loop order~\cite{alcantara_80}.

In the minimal renormalization procedure, i.e., dimensional regularization in conjunction with minimal subtraction, the $Z$-factors are of the form
\begin{equation}
\label{expZ}
Z_{\ldots }(u) =1+\sum_{m=1}^{\infty }\frac{X_{\ldots }^{(m)}(u)}{\varepsilon^{m}}\, .
\end{equation}
The $X_{\ldots }^{(m)}(u)$ are expansions in the coupling constant $u$ beginning with the power $u^{m}$. It is a fundamental fact of renormalization theory, cf.~Ref.~\cite{amit_zinn-justin}, that this procedure is suitable to eliminate the UV-divergencies from any vertex function order by order in perturbation theory. 

The unrenormalized theory has to be independent of the arbitrary length scale $\mu^{-1}$ introduced by renormalization. Hence, the unrenormalized correlation functions satisfy the identity
\begin{eqnarray}
\label{independence}
\mu \frac{\partial}{\partial \mu} \mathaccent"7017{G}_N \left( \left\{ {\rm{\bf x}} ,\vec{\lambda} \right\} ;  \mathaccent"7017{g}, \mathaccent"7017{\tau}, \mathaccent"7017{w} \right) = 0 \, . 
\end{eqnarray}
Equation~(\ref{independence}) translates via the Wilson functions
\begin{subequations}
\label{wilson}
\begin{eqnarray}
\label{wilsonGamma}
\gamma_{\ldots } \left( u \right) &=& \mu \frac{\partial }{\partial \mu} \ln Z_{\ldots } \bigg|_0 \, ,
\\
\label{betau}
\beta \left( u \right) &=& \mu \frac{\partial u}{\partial \mu} \bigg|_0 = u \left( 3 \gamma - \gamma_u - \varepsilon \right) \, ,
\\
\kappa \left( u \right) &=& \mu \frac{\partial \ln \tau}{\partial \mu} \bigg|_0 = \gamma - \gamma_\tau \, ,
\\
\label{wilsonZeta}
\zeta \left( u \right) &=& \mu \frac{\partial \ln w}{\partial \mu} \bigg|_0 = \gamma - \gamma_w \, ,
\end{eqnarray}
\end{subequations}
(the $|_0$ indicates that bare quantities are kept fixed while taking the derivatives) into the RGE
\begin{eqnarray}
&&\left[ \mu \frac{\partial }{\partial \mu} + \beta \frac{\partial }{\partial u} + \tau \kappa \frac{\partial }{\partial \tau} + w \zeta 
\frac{\partial }{\partial w} + \frac{N}{2} \gamma \right] 
\nonumber \\
&&\times \, 
G_N \left( \left\{ {\rm{\bf x}} ,\vec{\lambda} \right\} ; u, \tau, w, \mu \right) = 0 \, .
\end{eqnarray}

From the structure of the renormalization factors (\ref{completeZfactors}) and the Wilson functions (\ref{wilson}) one can deduce the important fact the RGE is determined entirely by the $X_{\ldots }^{(1)}(u)$. This may be seen as follows. From the definitions (\ref{wilsonGamma}) and (\ref{betau}) we learn that the Wilson $\gamma$ functions can be expressed as
\begin{eqnarray}
\label{rrrrr}
\gamma_{\ldots } \left( u \right) = \beta \left( u \right)\frac{\partial }{\partial u} \ln Z_{\ldots } \, .
\end{eqnarray}
A glance at Eq.~(\ref{expZ}) tells us then that the logarithmic derivative in Eq.~(\ref{rrrrr}) has a pure Laurent expansion with respect to $\varepsilon$ starting at first order in $1/\varepsilon$. Moreover, we know from Eq.~(\ref{betau}) that $\beta (u)$ begins with the zero-loop term $-\varepsilon u$. Because the $\gamma$ functions are finite for $\varepsilon \to 0$ their $\varepsilon$ poles have to cancel order by order in the loop expansion. As a consequence, the $\gamma$ functions are given by
\begin{eqnarray}
\gamma_{\ldots } \left( u \right) = -u \frac{\partial }{\partial u} X_{\ldots }^{(1)}(u) \, .
\end{eqnarray}
For this reason we will focus in the remainder of this paper on the $X_{\ldots }^{(1)}(u)$. We neglect higher order terms in the expansion~(\ref{expZ}) and write
\begin{equation}
\label{completeZfactors}
Z_{\ldots }(u) =1+ \sum_{L=1}^{\infty }\frac{Y^{(L)}_{\ldots }}{L \, \varepsilon} \, u^L + O \left( \varepsilon^{-2} \right) \, ,
\end{equation}
where the $Y^{(L)}_{\ldots }$ are numerical coefficients independent of $\varepsilon$. 

The RGE can be solved in terms of a single flow parameter $l$ by using the characteristics
\begin{subequations}
\begin{eqnarray}
l \frac{\partial \bar{\mu}}{\partial l} &=& \bar{\mu} \, , \quad \bar{\mu}(1)=\mu \ ,
 \\
\label{charBeta}
l \frac{\partial \bar{u}}{\partial l} &=& \beta \left( \bar{u}(l) \right) \, , \quad \bar{u}(1)=u \, ,
 \\
l \frac{\partial}{\partial l} \ln \bar{\tau} &=& \kappa \left( \bar{u}(l) \right) \, , \quad \bar{\tau}(1)=\tau \, ,
 \\
l \frac{\partial}{\partial l} \ln \bar{w} &=& \zeta \left( \bar{u}(l) \right) \, , \quad \bar{w}(1)=w \, ,
 \\
l \frac{\partial}{\partial l} \ln \bar{Z} &=& \gamma \left( \bar{u}(l) \right) \, , \quad \bar{Z}(1)=1 \, .
\end{eqnarray}
\end{subequations}
These characteristics describe how the parameters transform if we change the momentum scale $\mu $ according to $\mu \to \bar{\mu}(l)=l\mu $. Being interested in the infrared (IR) behavior of the theory, we study the limit $l\to 0$. According to Eq.~(\ref{charBeta}) we expect that in this IR limit the coupling constant $\bar{u}(l)$ flows to a stable fixed point $u^\ast$ satisfying $\beta (u^\ast )=0$. The IR stable fixed point solution to the RGE is readily found. In conjunction with dimensional analysis it gives 
\begin{eqnarray}
\label{scaling}
&&G_N \left( \left\{ {\rm{\bf x}} ,\vec{\lambda} \right\} ;  u, \tau , w, \mu \right) = 
l^{(d-2+\eta)N/2} 
\nonumber \\
&& \times \,
G_N \left( \left\{ l{\rm{\bf x}} , \vec{\lambda} \right\} ;  u^\ast, l^{-1/\nu}\tau , l^{-\phi /\nu} w, \mu \right)
\nonumber \\
\end{eqnarray}
with the critical exponents for percolation $\eta = \gamma (u^\ast )$ and $\nu = [2-\kappa (u^\ast)]^{-1}$ known to third order in $\varepsilon$~\cite{alcantara_80}. $\phi = \nu [2- \zeta (u^\ast )]$ is the percolation resistance exponent known to second order in $\varepsilon$~\cite{lubensky_wang_85,stenull_janssen_oerding_99,stenull_2000}.

Equation~(\ref{scaling}) implies that the two-point function $\widetilde{G} ( {\rm{\bf x}}, {\rm{\bf x}}^\prime , \vec{\theta} )$ scales at criticality $\tau =0$ as
\begin{eqnarray}
\label{scaling2punkt1}
\widetilde{G} \left( {\rm{\bf x}}, {\rm{\bf x}}^\prime , \vec{\theta} \right) = l^{2\beta /\nu} \Omega \left( l \left| {\rm{\bf x}} - {\rm{\bf x}}^\prime \right|, l^{-\phi /\nu} w^{-1} \vec{\theta}^2 \right)
\nonumber \\
\end{eqnarray}
with $\beta = (d-2+\eta) \nu /2$ denoting the percolation order parameter exponent and where $\Omega$ is a scaling function. Upon choosing $l = | {\rm{\bf x}} - {\rm{\bf x}}^\prime |^{-1}$ and expanding the right hand side of Eq.~(\ref{scaling2punkt1}) we obtain by dropping non-universal constants
\begin{eqnarray}
\label{scaling2punkt2}
&&\widetilde{G} \left( {\rm{\bf x}}, {\rm{\bf x}}^\prime , \vec{\theta} \right) = \left| {\rm{\bf x}} - {\rm{\bf x}}^\prime \right|^{2\beta /\nu} 
\nonumber \\
&& \times \,
\bigg\{ -w^{-1} \frac{\vec{\theta}^2}{2} \left| {\rm{\bf x}} - {\rm{\bf x}}^\prime \right|^{ -\phi /\nu} + \cdots \bigg\} \, .
\end{eqnarray}
With help of Eq.~(\ref{melkeGenFkt}) it is now straightforward to deduce the scaling behavior of the average conductance,
\begin{eqnarray}
M_{R^{-1}} \sim w^{-1} \left| {\rm{\bf x}} - {\rm{\bf x}}^\prime \right|^{ -\phi /\nu}  \, .
\end{eqnarray}

\section{Renormalization group analysis of the \rsc}
\label{RGA_RSC}
Now we turn to the \rsc and assume that $0<a<1$. First we address the renormalization of the model. By carefully analyzing the RG flow we reveal the critical behavior of $M_{R^{-1}}^{\text{SC}}$ and $\Sigma^{\text{SC}}$. As far as notation is concerned, we adopt the same convention as in Secs.~\ref{intro} and \ref{theModel}. Quantities that might be confused with their analogs for the RRN are marked by the superscript SC.

\subsection{Renormalization}
\label{renormalization}
Obviously the Potts model $Z$ factors are independent of $a$ and hence do not require further consideration. The $Z$ factor pertaining to $w$, however, is expected to be different from its analog for the RRN, i.e., 
\begin{eqnarray}
\label{renoW}
w \to \mathaccent"7017{w} &=& Z^{-1} Z_w^{\text{SC}} w \, .
\end{eqnarray}
Since $v\neq 0$ we need an additional renormalization
\begin{eqnarray}
\label{renoV}
v \to \mathaccent"7017{v} &=& Z^{-1} Z_v^{\text{SC}} v \, .
\end{eqnarray}
From Eqs.~(\ref{renoW}) and (\ref{renoV}) we deduce immediately, that the effective invariant coupling $h = v/w^{1-a}$ announced in Sec.~\ref{scalingAnalysisVoltageVariables} has to be renormalized by
\begin{eqnarray}
\label{renoH}
h \to \mathaccent"7017{h} &=& Z^{-a} Z_v^{\text{SC}} \left( Z_w^{\text{SC}} \right)^{a-1} h \, \mu^{2a} \, .
\end{eqnarray}
The factor $\mu^{2a}$ is included to render the renormalized effective coupling dimensionless.

After these remarks we now bring our attention to the perturbation calculation and its Feynman diagrams. For the \rsc the principal propagator has a form similar to its analog for the RRN,
\begin{eqnarray}
G^{\text{SC},\text{bold}} ( \brm{p}, \vec{\lambda} ) = G^{\text{SC}} ( \brm{k}, \vec{\lambda} ) \{ 1 - \delta_{\vec{\lambda}, \vec{0}} \} \, .
\end{eqnarray}
However, its flesh is now given by $G^{\text{SC}} (\brm{p}, \vec{\lambda}) = (\tau + \brm{p}^2 + w \vec{\lambda}^2 + v \vec{\lambda}^{2(1-a)})^{-1}$. Evidently, $G^{\text{SC},\text{bold}} ( \brm{p}, \vec{\lambda} )$ decomposes into a conducting and an insulating part. This leads to conducting diagrams identical to those for the RRN up to apparent distinctions in the definition of the propagators. Due to these distinctions an expansion for small external momenta and currents leads to
\begin{eqnarray}
\label{diagrStructRSC}
I \left( {\rm{\bf p}}^2 , \vec{\lambda}^2 \right) &=& I_P \left( {\rm{\bf p}}^2 \right) - I_W \left( {\rm{\bf p}}^2 \right) w \vec{\lambda}^2 
\nonumber \\
&-& I_V \left( {\rm{\bf p}}^2 \right) v \vec{\lambda}^{2(1-a)} + \cdots
\end{eqnarray}
instead of Eq.~(\ref{diagrStructRRN}) for the overall form of the conducting diagrams.

To extract information on $Z_v^{\text{SC}}$ consider the $\vec{\lambda}$ dependent part of a conducting diagram with $P^{\text{cond}}$ conducting propagators which reads in Schwinger parameterization
\begin{eqnarray}
\label{lambdaPart1}
\sum_{\{ \vec{\kappa} \}} \exp \bigg[ - \sum_{i\in P^{\text{cond}}} s_i \Big( w \vec{\lambda}^2_i + v  \vec{\lambda}^{2(1-a)}_i \Big) \bigg]  \, .
\end{eqnarray}
We keep in mind that $\vec{\lambda}_i = \vec{\lambda}_i ( \vec{\lambda} , \{ \vec{\kappa} \} )$. The important observation is now that any conducting propagator affected by the summation over the loop currents gives a contribution to Eq.~(\ref{diagrStructRSC}) polynomial in $\vec{\lambda}$, i.e., it contributes to $I_W$ rather than to $I_V$. The only contributions to $I_V$ can come from conducting propagators not affected by the summation over $\{ \vec{\kappa} \}$. In the terminology of our real-world interpretation this means that $I_V$ is determined exclusively by the red bonds of that diagram, i.e., by its singly connected conducting propagators. To be specific, $I_V$ is given by
\begin{eqnarray}
\label{expansionOfDiagrams_v}
I_V \left( {\rm{\bf p}}^2 \right) = \int_0^\infty \prod_j ds_j \sum_{i\in P^{\text{red}}} 
s_i \, D \left( {\rm{\bf p}}^2, \left\{ s_j \right\} \right) \, ,
\end{eqnarray}
where the sum runs over all $P^{\text{red}}$ conducting propagators of the diagram not belonging to any closed conducting loop.

Now we take a short detour and recall some central features of our field theory on the nonlinear RRN~\cite{janssen_stenull_oerding_99&janssen_stenull_99,stenull_2000} in which the occupied bonds obey a generalized Ohm's law $V\sim I^r$. The field theoretic Hamiltonian for the nonlinear RRN~\cite{harris_87} corresponds to that for the standard RRN with $w \nabla_{\vec{\theta}}^2$ replaced by $w_r \sum_{\alpha =1}^D (-\partial / \partial \theta^{(\alpha )})^{r+1}$. Accordingly, one encounters a generalized renormalization factor $Z_{w_r}$ that then leads to a generalized resistance exponent $\phi_r = \nu [2-\zeta_r (u^\ast )]$. The Wilson function $\zeta_r$ is defined analogous to Eq.~(\ref{wilsonZeta}) with $\gamma_w$ replaced by $\gamma_{w_r} = \mu \frac{\partial}{\partial \mu} \ln Z_{w_r} |_0$. The generalized resistance exponent has the physical meaning of governing the average nonlinear resistance at criticality,
\begin{eqnarray}
\label{nonLinRes}
M_r (x ,x^\prime) &=& \left\langle \chi (x ,x^\prime) R_r (x ,x^\prime )  \right\rangle_C 
/ \left\langle \chi (x ,x^\prime) \right\rangle_C  
\nonumber \\
&\sim& \left| x - x^\prime \right|^{\phi_r/\nu} \, .
\end{eqnarray}

The nonlinear RRN is particularly interesting, because it provides for an elegant way to determine fractal dimension of percolation clusters by considering specific values of $r$. For example, it is a well known fact~\cite{blumenfeld_aharony_85} that
\begin{eqnarray}
\label{relRed}
\lim_{r\to \infty }M_r \sim M_{\text{red}}  \, ,
\end{eqnarray}
where $M_{\text{red}}$ is the mass (average number) of the red bonds. From Eqs.~(\ref{nonLinRes}) and (\ref{relRed}) one obtains immediately the scaling relation $d_{\text{red}} = \lim_{r\to \infty} \phi_r /\nu$ for the fractal dimension $d_{\text{red}}$ of the red bonds. It was shown rigorously by Coniglio~\cite{coniglio_81_82} that $d_{\text{red}} = 1/\nu$ which means that $\lim_{r\to \infty} \phi_r =1$. From the definition of $\phi_r$ it then follows that $\lim_{r\to \infty}\zeta_r (u) = \kappa (u)$ which leads in turn to the identity
\begin{eqnarray}
\label{ward1}
\lim_{r\to \infty } Z_{w_r} = Z_\tau \, .
\end{eqnarray}
In Refs.~\cite{janssen_stenull_oerding_99&janssen_stenull_99,stenull_2000} we showed based on our real-world interpretation that the contribution $I_{W_r}$ of a conducting diagram to $Z_{w_r}$ takes for $r\to \infty$ the simple form
\begin{eqnarray}
\label{expansionOfDiagrams_r_infty}
\lim_{r\to \infty } I_{W_r} \left( \brm{p}^2 \right) = \int_0^\infty \prod_j ds_j \sum_{i\in P^{\text{red}}} s_i \, D \left( {\rm{\bf p}}^2, \left\{ s_j \right\} \right) \, .
\nonumber \\
\end{eqnarray}
Comparison of Eqs.~(\ref{expansionOfDiagrams_v}) and (\ref{expansionOfDiagrams_r_infty}) yields
\begin{eqnarray}
\label{ward2}
\lim_{r\to \infty } Z_{w_r} = Z_v^{\text{SC}}\, .
\end{eqnarray}
From this and Eq.~(\ref{ward1}) we finally conclude the identity
\begin{eqnarray}
\label{ward3}
Z_v^{\text{SC}} = Z_\tau \, .
\end{eqnarray}

For analyzing $Z_w^{\text{SC}}$ we revisit Eq.~(\ref{lambdaPart1}) and rescale the replica currents, $\vec{\lambda}^2 \to w^{-1} \vec{\lambda}^2$. This recasts Eq.~(\ref{lambdaPart1}) into
\begin{eqnarray}
\label{lambdaPart2}
\sum_{\{ \vec{\kappa} \}} \exp \bigg[ - \sum_{i\in P^{\text{cond}}} s_i \Big( \vec{\lambda}^2_i + h  \vec{\lambda}^{2(1-a)}_i \Big) \bigg]  \, .
\end{eqnarray}
As mentioned above, conducting propagators belonging to closed conducting loops (blobs) lead in an expansion for small $\vec{\lambda}$ to terms polynomial in $\vec{\lambda}$ that contribute to $I_W$. Since $h$ appears in  Eq.~(\ref{lambdaPart2}) $I_W$ will in general depend on $h$ (cf.~Appendix~\ref{app:YYY}). We conclude that $Z_w^{\text{SC}}$ is not only a function of $u$ but also of $h$,
\begin{eqnarray}
Z_w^{\text{SC}} = Z_w^{\text{SC}} (u,h) \, .
\end{eqnarray}

For arbitrary $a\in (0,1)$ it is difficult to gain further insight into $Z_w^{\text{SC}}$. Anyway, the most exciting values of $a$ are those for which $a$ is of the order of our small expansion parameter $\varepsilon$. This is the key region in which the crossover between the RRN and the \rsc occurs (the naive limit $a\to 0$ presupposes $\varepsilon \ll a$ and hence is inadequate to resolve the crossover). In case both $a$ and $\varepsilon$ are small we can follow the work of Honkonen and Nalimov~\cite{honkonen_nalimov_89} to analyze the structure of $Z_w^{\text{SC}}$. This structure differs from that of $Z_w$ [cf.\ Eq.~(\ref{completeZfactors})] because two major differences emerge for $a>0$. First, the conducting propagators
\begin{eqnarray}
G^{\text{SC},\text{cond}} ( \brm{p}, w^{-1/2}\vec{\lambda} ) = \sum_{k=0}^\infty \frac{(-1)^k \, \vec{\lambda}^{2(1-a)k}}{\big[ \tau + \brm{p}^2 + \vec{\lambda}^2 \big]^{k+1}} \, h^k
\nonumber \\
\end{eqnarray}
give rise to an entire series of individual terms. Second, since $a$ is now of order $\varepsilon$, the poles in these individual contributions are now of the type $1/(L\varepsilon + 2ka)$. Hence $Z_w^{\text{SC}}$ is of the form
\begin{eqnarray}
\label{formZWSC}
Z_w^{\text{SC}} (u,h) = 1 + \sum_{L=1}^\infty \sum_{k=0}^\infty  \frac{Y^{(L,k)}_w}{L\, \varepsilon + 2 k  a} \, u^l h^k + O \left( \varepsilon^{-2} \right) \, .
\nonumber \\
\end{eqnarray}
Here, the $Y^{(L,k)}_w$ are the numerical coefficients of the poles. Appendix~\ref{app:YYY} illustrates the preceding arguments at the instance of an one-loop calculation.

In minimal subtraction, the $Y^{(L,k)}_w$ are independent of both $\varepsilon$ and $a$. This fact provides us with a simple method to calculate the numerical coefficients because it is sufficient to consider the limit $a\to 0$. In this limit the Hamiltonian for the \rsc reduces to the Hamiltonian for the RRN with $w$ replaced by $w+v$. As a consequence we obtain by exploiting Eq.~(\ref{ward3}) the relation
\begin{eqnarray}
\label{blub}
Z_w^{\text{SC}} (u,h) = Z_w (u) + h \left[ Z_w (u) + Z_\tau (u) \right] 
\end{eqnarray}
valid in the limit $a\to 0$. Inserting the explicit forms~(\ref{completeZfactors}) and (\ref{formZWSC}) into Eq.~(\ref{blub}) we obtain the following relations between the numerical coefficients: $Y^{(L,0)}_w = Y^{(L)}_w$, $Y^{(L,1)}_w = Y^{(L)}_w-Y^{(L)}_\tau$, and $Y^{(L,k>1)}_w = 0$. Recalling that the $Y^{(L,k)}_w$ are independent of $a$ we conclude that
\begin{eqnarray}
\label{resZWSC}
Z_w^{\text{SC}} (u,h) &=& 1 + \sum_{L=1}^\infty  u^l \Bigg\{  \frac{Y^{(L)}_w}{L\, \varepsilon} + h \, \frac{Y^{(L)}_w + Y^{(L)}_\tau}{L\, \varepsilon + 2\, a} \Bigg\} 
\nonumber \\
&+& O \left( \varepsilon^{-2} \right) \, .
\end{eqnarray}

\subsection{Scaling}
\label{scalingRSC}
We proceed in the same fashion as in Sec.~\ref{RGA_RRN} and set up a Gell--Mann-Low RG equation for the \rsc. By carefully analyzing the RG flow we derive the scaling behavior of the average conductance. 

The RGE for the \rsc is somewhat richer than that for the RRN:
\begin{eqnarray}
\label{RGE_RSC}
&&\left[ \mu \frac{\partial }{\partial \mu} + \beta \frac{\partial }{\partial u} + \tau \kappa \frac{\partial }{\partial \tau} + w \zeta_w^{\text{SC}} \frac{\partial }{\partial w} + v \zeta_v^{\text{SC}} \frac{\partial }{\partial v}+ \frac{N}{2} \gamma \right] 
\nonumber \\
&&\times \, 
G_N \left( \left\{ {\rm{\bf x}} ,\vec{\lambda} \right\} ; u, \tau, w, v, \mu \right) = 0 \, ,
\end{eqnarray}
where we have introduced the Wilson functions
\begin{subequations}
\label{wilsonRSC}
\begin{eqnarray}
\label{gammaW}
\gamma_w^{\text{SC}} \left( u ,h\right) &=& \mu \frac{\partial }{\partial \mu} \ln Z_{w}^{\text{SC}} \bigg|_0 \, ,
\\
\label{zetaW}
\zeta_w^{\text{SC}} \left( u,h \right) &=& \mu \frac{\partial \ln w}{\partial \mu} \bigg|_0 = \gamma - \gamma_w^{\text{SC}} \, ,
\\
\label{zetaV}
\zeta_v^{\text{SC}} \left( u \right) &=& \mu \frac{\partial \ln v}{\partial \mu} \bigg|_0 = \gamma - \gamma_v^{\text{SC}} \, .
\end{eqnarray}
\end{subequations}
Obviously value of $\zeta_v^{\text{SC}}$ at the fixed point $u^\ast$ is given by 
\begin{eqnarray}
\label{fixedV}
\zeta_v^{\text{SC}} \left( u^\ast \right) = 2 - 1/\nu \, . 
\end{eqnarray}
In order to express $\zeta_w^{\text{SC}}$ at $u^\ast$ in terms of the known RRN exponents we introduce the function
\begin{eqnarray}
\label{definitionOf_f}
f(h) = \gamma_w^{\text{SC}} \left( u^\ast ,h\right) - \gamma_w \left( u^\ast \right)\, ,
\end{eqnarray}
which leads to 
\begin{eqnarray}
\label{fixedW}
\zeta_w^{\text{SC}} \left( u^\ast ,h \right)= 2 - \phi /\nu - f(h) \, .
\end{eqnarray}
Equations~(\ref{wilsonRSC}), (\ref{fixedV}), and (\ref{fixedW}) provide us with flow equations for the couplings $w$ and $v$:
\begin{subequations}
\begin{eqnarray}
\label{flowW}
l \frac{\partial}{\partial l} \ln \bar{w} (l)&=& 2 - \phi /\nu - f\left( \bar{h} (l) \right) \, , \quad \bar{w}(1)=w \, ,
\nonumber \\
\\
\label{flowV}
l \frac{\partial}{\partial l} \ln \bar{v}(l) &=& 2 - 1 /\nu  \, , \quad \bar{v}(1)=v \, .
\end{eqnarray}
\end{subequations}
The flow equation~(\ref{flowV}) for $v$ is readily solved,
\begin{eqnarray}
\label{solV}
\bar{v} (l) = v \, l^{2-1/\nu} \, .
\end{eqnarray}
To solve the flow equation~(\ref{flowW}) for $w$ we first have to analyze the flow of $h$. 

From the renormalization of $h$ (\ref{renoH}) follows immediately that the logarithmic derivative of the renormalized $h$ with respect to $\mu$ is given at $u^\ast$ by
\begin{eqnarray}
\mu \frac{\partial}{\partial \mu} \ln h \bigg|_0 \stackrel{u=u^\ast}{=} \frac{1}{\nu} \Big[ (1-a)\phi -1 \Big] + (1-a) f(h) \, .
\nonumber \\
\end{eqnarray}
Consequently, $h$ obeys the flow equation
\begin{eqnarray}
\label{flowH}
l \frac{\partial}{\partial l} \bar{h} (l) &=& \bar{h} (l) \, \bigg\{ \frac{1}{\nu} \Big[ (1-a)\phi -1 \Big] + (1-a) f\left( \bar{h} (l) \right) \bigg\}
\nonumber \\
\end{eqnarray}
with the initial condition $\bar{h}(1)=h$. A glance reveals that the flow of $h$ has two IR fixed points, viz.\ $h^\ast_1 =0$ and $h^\ast_2$ determined by
\begin{eqnarray}
\label{condH2}
f \left( h^\ast_2 \right) = \frac{1}{\nu} \bigg[ \frac{1}{1-a} - \phi \bigg] \, .
\end{eqnarray}

Now that we know the fixed points of $h$ we have to analyze their stability. Since there are two fixed points either one will be stable and the other one will be unstable depending on the value of $a$. Consider $h^\ast_1$. $h^\ast_1$ is stable if the $\{ \cdots \}$ bracket on the right hand side of Eq.~(\ref{flowH}) has a positive sign and it is unstable if this bracket has a negative sign. For $a$ of order one the sign is certainly negative. This leads to the conclusion that $h^\ast_2$ is stable for $a$ of order one. In the crossover region, i.e., for $a$ of order $\varepsilon$, the question of stability is more intricate. Here, we need more information on the functional dependence of $f$ on $h$. This information can be derived from the structure of $Z_w^{\text{SC}}$ given in Eq.~(\ref{resZWSC}). Upon inserting Eq.~(\ref{resZWSC}) into the definition~(\ref{gammaW}) of $\gamma_w^{\text{SC}}$ we obtain 
\begin{eqnarray}
\label{finalGammaW}
\gamma_w^{\text{SC}} \left( u ,h\right) = \sum_{L=1}^\infty  u^l \Big\{ Y^{(L)}_w + h \left[ Y^{(L)}_w - Y^{(L)}_\tau \right] \Big\} \, .
\nonumber \\
\end{eqnarray}
Substituting $\gamma_w^{\text{SC}} (u^\ast ,h)$ into the definition (\ref{definitionOf_f}) of $f$ leads to
\begin{eqnarray}
\label{finalF}
f(h) = \frac{h}{\nu} \left[ \phi - 1 \right] \, .
\end{eqnarray}
The important conclusion from Eq.~(\ref{finalF}) is that $f$ is linear in $h$ for $a$ of order $\varepsilon$. Hence, it results in a contribution to the right hand side of Eq.~(\ref{flowH}) quadratic in $h$ that can be neglected in analyzing the stability of $h^\ast_1$. Finally, we perceive that $h^\ast_1$ is stable if $\phi > (1-a)^{-1}$ whereas $h^\ast_2$ is stable if $\phi < (1-a)^{-1}$.

At this point we possess enough information to solve the flow equation~(\ref{flowW}) for $w$. For $\phi > (1-a)^{-1}$ we know that $f$ tends to zero because $h$ flows to $h^\ast_1$. For $\phi < (1-a)^{-1}$, on the other hand, we have to insert (\ref{condH2}). Summarizing both cases we write the solution to Eq.~(\ref{flowW}) as
\begin{eqnarray}
\label{solW}
\bar{w} (l) = w \, l^{2 -\phi^{\text{SC}} (a) /\nu} \, ,
\end{eqnarray}
where we have defined the $a$ dependent resistance exponent 
\begin{eqnarray}
\label{phiVonA}
\phi^{\text{SC}} (a) =
\left\{ 
\begin{array}{ccc}
\phi &\mbox{if}& \phi > \frac{1}{1-a} \\
\frac{1}{1-a} &\mbox{if}& \phi < \frac{1}{1-a}
\end{array}
\right. \, .
\end{eqnarray}

Collecting the above results we find that the solution to the RGE~(\ref{RGE_RSC}) augmented by dimensional analysis reads
\begin{eqnarray}
\label{scalingRSC1}
&&G_N \left( \left\{ {\rm{\bf x}} ,\vec{\lambda} \right\} ;  u, \tau , w, v, \mu \right) = 
l^{N\beta /\nu} 
\nonumber \\
&& \times \,
G_N \left( \left\{ l{\rm{\bf x}} , \vec{\lambda} \right\} ;  u^\ast, l^{-1/\nu}\tau , l^{-\phi^{\text{SC}} (a) /\nu} w, l^{-1/\nu} v ,\mu \right) \, .
\nonumber \\
\end{eqnarray}
This scaling form implies for the Fourier transformed two-point function at criticality that
\begin{eqnarray}
\label{scaling2punktSC1}
&&\widetilde{G} \left( {\rm{\bf x}}, {\rm{\bf x}}^\prime , \vec{\theta} \right) = l^{2\beta /\nu} 
\nonumber \\
&& \times \,
\Xi \left( l \left| {\rm{\bf x}} - {\rm{\bf x}}^\prime \right|, l^{-\phi^{\text{SC}} (a) /\nu} w^{-1} \vec{\theta}^2 , h^\ast_{1,2}\right) \, ,
\end{eqnarray}
where $\Xi$ is a scaling function. The choice $l = | {\rm{\bf x}} - {\rm{\bf x}}^\prime |^{-1}$ and subsequent Taylor expansion yields
\begin{eqnarray}
\label{scaling2punktSC2}
&&\widetilde{G} \left( {\rm{\bf x}}, {\rm{\bf x}}^\prime , \vec{\theta} \right) = \left| {\rm{\bf x}} - {\rm{\bf x}}^\prime \right|^{2\beta /\nu} 
\nonumber \\
&& \times \,
\bigg\{ -w^{-1} \frac{\vec{\theta}^2}{2} \left| {\rm{\bf x}} - {\rm{\bf x}}^\prime \right|^{ -\phi^{\text{SC}} (a) /\nu} + \cdots \bigg\} \, .
\end{eqnarray}
With help of Eq.~(\ref{melkeGenFkt}) we readily obtain
\begin{eqnarray}
\label{scaleFormConductance1}
M^{\text{SC}}_{R^{-1}} \sim w^{-1} \left| {\rm{\bf x}} - {\rm{\bf x}}^\prime \right|^{ -\phi^{\text{SC}} (a) /\nu}  \, .
\end{eqnarray}

It remains to deduce the scaling behavior the average conductivity and its conductivity exponent $t^{\text{SC}} (a)$. Commonly, the conductivity of percolating systems is defined with respect to a bus bar geometry where the network is placed between two parallel superconducting plates (the electrodes) of area $L^{d-1}$ a distance $L$ apart. From the above we expect that the average conductance $\sigma^{\text{SC}}$ of this system scales as
\begin{eqnarray}
\label{scaleFormConductance2}
\sigma^{\text{SC}} (L, \tau) = \left| \tau \right|^{ \phi^{\text{SC}} (a)} \Pi_a \left( L/\xi \right)  \, ,
\end{eqnarray}
where $\Pi_a$ is an $a$-dependent scaling function with the properties
\begin{eqnarray}
\label{propsPi}
\Pi_a (x) \sim
\left\{ 
\begin{array}{ccc}
const &\mbox{for}& x \ll 1 \\
x^{d-2} &\mbox{for}& x \gg 1
\end{array}
\right. \, .
\end{eqnarray}
Now consider length scales large compared to the correlation length $\xi$. In this regime the \rsc above the percolation threshold, $\tau <0$, can be viewed as a homogeneous system of conductivity $\Sigma^{\text{SC}} (\tau)$. Hence, we may write for $L\gg \xi$ that
\begin{eqnarray}
\Sigma^{\text{SC}} (\tau) \sim L^{2-d} \sigma^{\text{SC}} (L, \tau) \, .
\end{eqnarray}
By virtue of Eq.~(\ref{propsPi}) we finally get
\begin{eqnarray}
\Sigma^{\text{SC}} (\tau) \sim  \left| \tau \right|^{ (d-2)\nu + \phi^{\text{SC}} (a)} \, ,
\end{eqnarray}
which means that the conductivity exponent is given by
\begin{eqnarray}
t^{\text{SC}} (a) = (d-2)\nu + \phi^{\text{SC}} (a) \, .
\end{eqnarray}

\section{Concluding remarks}
\label{concusions} 
We showed without relying on the assumptions of the 'nodes, links, and blobs' picture that the conductivity exponent for the \rsc is given by Eq.~(\ref{tVonA}). For sufficiently large values of $a$ this means that
\begin{eqnarray}
t^{\text{SC}} (a) = (d-2)\nu + (1-a)^{-1} \, .
\end{eqnarray}
recognized by Halperin {\em et al}.~\cite{halperin&co_85_87} as a lower bound to $t^{\text{SC}} (a)$ that can be attributed to a dominance of the red bonds. At a critical value $a_c = 1 -1/\phi$ the conductivity exponent crosses over to the value
\begin{eqnarray}
t^{\text{SC}} (a) = (d-2)\nu + \phi 
\end{eqnarray}
for the standard lattice model and is essentially determined by the blobs.

The 'nodes, links, and blobs' picture provides an intuitive explanation for this behavior. Since the distribution function (\ref{fDef}) is singular, there
exists at criticality a large number of bonds with arbitrarily large resistance. Whenever one of these bonds is red it may dominate the total resistance of its link. If the large resistance, on the other hand, occurs in a blob it has, in general, little impact because the current can flow through parallel paths. For $a>a_c$ the resistance of the weakest red bond is typically larger than the sum of resistances of the other building blocks (the blobs and the other red bonds) of a link so that the weakest red bond dominates the total resistance of its entire link. For $a<a_c$ the importance of the weak red bonds is diminished and the network behaves effectively as a standard RRN.

We point out that our analysis was not restricted to any particular order in $\varepsilon$-expansion. Unlike LT who set up a 'momentum shell' RG, we used the powerful methods of renormalized field theory. These methods allowed us to explore general properties of the renormalization factors, subsequently the RGE, and finally the conductivity exponent to all orders in perturbation theory.

\begin{acknowledgments}
We acknowledge the support by the Sonderforschungsbereich 237 ``Unordnung und gro{\ss}e Fluktuationen'' of the Deutsche Forschungsgemeinschaft. 
\end{acknowledgments}  

\appendix
\section{Contents of the coupling constants $w$ and $v$}
\label{app:XXX}
In this Appendix we give details on the integrations in Eq.~(\ref{vieleInts}). At the instance of $F_1$ and $F_2$ we illustrate the contents of the coupling constants $w$ and $v$.

We start by recalling the definition of $F_1$. After changing the integration variable from $\sigma$ to $t$ by setting $t=\vec{\lambda}^2/(2\sigma)$ $F_1$ takes on the form
\begin{eqnarray}
F_1 \left( \vec{\lambda} \right) = (1-a) \left( \frac{\vec{\lambda}^2}{2\sigma_0} \right)^{1-a} \Gamma \left( a-1, \frac{\vec{\lambda}^2}{2\sigma_0} \right) \, .
\end{eqnarray}
Here 
\begin{eqnarray}
\Gamma (\alpha ,y) = \int_y^\infty dt \, t^{\alpha -1} e^{-t}
\end{eqnarray}
is the incomplete Gamma function that has about $y=0$ the Taylor expansion~\cite{gradshteyn_ryzhik}
\begin{eqnarray}
\label{expGamma}
\Gamma (\alpha ,y) = \Gamma (\alpha ) - \sum_{n=0}^\infty \frac{(-1)^n y^{\alpha +n}}{n! \,  (\alpha +n)} \, .
\end{eqnarray}
Thus, we obtain
\begin{eqnarray}
\label{expF1}
F_1 \left( \vec{\lambda} \right) &=& 1 - \Gamma (a) \left( \frac{\vec{\lambda}^2}{2\sigma_0} \right)^{1-a} + \frac{1-a}{a} \frac{\vec{\lambda}^2}{2\sigma_0} 
\nonumber \\
&&+\, O \Big( \big( \vec{\lambda}^2 \big)^2 \Big) \, .
\end{eqnarray}

Now we turn to $F_2$. After isolating the contribution for $\vec{\lambda} =\vec{0}$ it is save to change variables by setting $\sigma_1 = \sigma_0 tx$ and $\sigma_2 = \sigma_0 t(1-x)$. We get
\begin{eqnarray}
F_2 \left( \vec{\lambda} \right) &=& 1 + (1-a)^2 \int_0^2 dt \int_0^1 dx \, t^{1-2a} x^{-a} (1-x)^{-a} 
\nonumber \\
&& \times \, 
\bigg\{ \exp \bigg( -  \frac{\vec{\lambda}^2}{2\sigma_0 t} \bigg) -1 \bigg\} \, .
\end{eqnarray}
The integration over $x$ gives a Beta function~\cite{gradshteyn_ryzhik} $B (1-a, 1-a)$. The integration over $t$ can be handled analogous to the integration in the preceding paragraph. As a result of both integrations we obtain
\begin{eqnarray}
F_2 \left( \vec{\lambda} \right) &=& 1 + (1-a)^2  B (1-a, 1-a)\, \Bigg\{ - \frac{2^{2(1-a)}}{2(1-a)}
\nonumber \\
&& + \, \left( \frac{\vec{\lambda}^2}{2\sigma_0} \right)^{2(1-a)} \Gamma \left( a-1, \frac{\vec{\lambda}^2}{4\sigma_0} \right) 
 \Bigg\} \, .
\end{eqnarray}
With help of the expansion~(\ref{expGamma}) we finally get
\begin{eqnarray}
F_2 \left( \vec{\lambda} \right) &=& 1 + (1-a)^2  B (1-a, 1-a)\, \Bigg\{ - \frac{2^{2(1-a)}}{2(1-a)}
\nonumber \\
&& \times \, 
\bigg\{ \Gamma \big( 2(a-1) \big) \left( \frac{\vec{\lambda}^2}{4\sigma_0} \right)^{2(1-a)} + \frac{1}{2(a-1)} \frac{\vec{\lambda}^2}{4\sigma_0} 
\nonumber \\
&&+\, O \Big( \big( \vec{\lambda}^2 \big)^2 \Big)
\Bigg\} \, .
\end{eqnarray}

The higher orders $F_{l>2}$ can be analyzed by similar means. It is not difficult to convince oneself that the general form of the $F_l$ is given by Eq.~(\ref{resF}) and that the coefficients $A_l$ decorating $\vec{\lambda}^2$ are proportional to $[1+l(a-1)]^{-1}$. Since the $F_l$ enter into the kernel $\widetilde{K} (\vec{\lambda})$ via Eq.~(\ref{expansionK}) the coefficient of the $\vec{\lambda}^2$-term in $\widetilde{K} (\vec{\lambda})$ is given by
\begin{eqnarray}
\label{tripleX}
w = \sum_{l=1}^\infty \frac{(-1)^l}{l} \, \upsilon^l A_l = \sum_{l=1}^\infty \frac{(-1)^l}{l} \, \upsilon^l \frac{C_l}{1+l(a-1)} \, ,
\end{eqnarray}
where the $C_l$ are positive constants proportional to $\sigma_0^{-1}$. We learn from Eq.~(\ref{tripleX}) that the value of $w$ depends on details like the specific values of $\upsilon$ and $a$. It may be finite or infinite, positive or negative. $v$ depends on $\upsilon$ and $a$ as well, see Eqs.~(\ref{expansionK}) and (\ref{expF1}). However, for $0<a<1$ and $p>0$ it is always finite and positive.

\section{One-loop calculation}
\label{app:YYY}
In this Appendix we calculate the one-loop contributions to $Z_w^{\text{SR}}$ explicitly. We will see that the one-loop result is in conformity with Eq.~(\ref{formZWSC}). 

\begin{figure}
\epsfxsize=8.7cm
\begin{center}
\epsffile{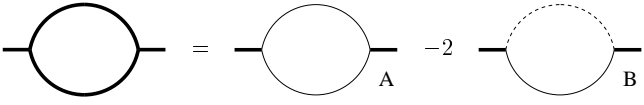}
\end{center}
\caption[]{\label{fig1}The principal one-loop self-energy diagram (bold) decomposes into the conducting diagrams A and B assembled of conducting (light) and insulating (dashed) propagators.}
\end{figure}

In one-loop order there exists only a single principal self-energy diagram that decomposes into conducting diagrams as shown in Fig.~\ref{fig1}.
In the following we set external momenta to zero for simplicity. Hence the one-loop contribution $\Gamma_2^{\text{1-loop}}$ to the vertex function $\Gamma_2$ (note that the vertex functions $\Gamma_N$ are defined as the negative of the corresponding diagrams) is given by
\begin{eqnarray}
\Gamma_2^{\text{1-loop}} \Big( \vec{\lambda} \Big) = 2\mbox{B} - \mbox{A} \, .
\end{eqnarray}
Applying the usual Feynman rules yields
\begin{eqnarray}
\label{step1}
\Gamma_2^{\text{1-loop}} \Big( \vec{\lambda} \Big) &=& - \frac{g^2}{2} \int_{\brm{q}} \sum_{\vec{\kappa}} G^{\text{cond}} \Big( \brm{q}, \vec{\kappa} \Big) G^{\text{cond}} \Big( \brm{q}, \vec{\kappa} + \vec{\lambda}\Big)
\nonumber \\
&& + \, g^2 \int_{\brm{q}} G^{\text{cond}} \Big( \brm{q}, \vec{\lambda}\Big) G^{\text{ins}} \Big( \brm{q}\Big)
\, ,
\end{eqnarray}
where we have dropped the superscript SC because we consider exclusively the \rsc throughout the entire Appendix. $\int_{\brm{q}}$ is an abbreviation for $(2\pi )^{-d} \int_{-\infty}^\infty d^d q$. For the following steps it is convenient to recast Eq.~(\ref{step1}) as
\begin{eqnarray}
\label{step2}
\Gamma_2^{\text{1-loop}} \Big( \vec{\lambda} \Big) &=& - \frac{g^2}{2} \int_{\brm{q}} \sum_{\vec{\kappa}} \bigg\{ G^{\text{cond}} \Big( \brm{q}, \vec{\kappa} \Big)^2 
\nonumber \\
&&
- \, \frac{1}{2} \Big[ G^{\text{cond}} \Big( \brm{q}, \vec{\kappa} + \vec{\lambda}\Big) - G^{\text{cond}} \Big( \brm{q}, \vec{\kappa} \Big) \Big]
\bigg\}
\nonumber \\
&& + \, g^2 \int_{\brm{q}} G^{\text{cond}} \Big( \brm{q}, \vec{\lambda}\Big) G^{\text{ins}} \Big( \brm{q}\Big)
\, .
\end{eqnarray}
The evaluation of the $\vec{\lambda}$-independent term is straightforward because $\sum_{\vec{\kappa}}G^{\text{cond}} ( \brm{q}, \vec{\kappa} )^2 = G^{\text{ins}} ( \brm{q})^2$ for $D\to 0$. The $\vec{\lambda}$-dependent terms are expanded in the external currents. After some algebra we obtain
\begin{eqnarray}
\label{step3}
&&\Gamma_2^{\text{1-loop}} \Big( \vec{\lambda} \Big) =  \frac{g^2}{2} \int_{\brm{q}} \frac{1}{\big( \tau + \brm{q}^2 \big)^2}
- g \int_{\brm{q}} \frac{w \vec{\lambda}^2 +v \vec{\lambda}^{2(1-a)}}{\big( \tau + \brm{q}^2 \big)^3}
\nonumber \\
&& + \, g^2 \int_{\brm{q}} \sum_{\vec{\kappa}} \frac{\big[w + (1-a) v \vec{\kappa}^{-2a} \big]^2 \big( \vec{\lambda} \cdot \vec{\kappa} \big)^2}{\big( \tau + \brm{q}^2 + w \vec{\kappa}^2 + v\vec{\kappa}^{2(1-a)}\big)^4} + \cdots
\, .
\end{eqnarray}
The first two integrations are readily carried out using dimensional regularization. The summation over the loop current can be simplified by exploiting the rotational symmetry in replica space and by rescaling $w \vec{\kappa}^2 \to \vec{\kappa}^2$. After these steps we arrive at
\begin{eqnarray}
\label{step4}
\Gamma_2^{\text{1-loop}} \Big( \vec{\lambda} \Big) &=&  - g^2 \frac{G_\varepsilon}{\varepsilon} \tau^{-\varepsilon /2} \Big\{ \tau + w \vec{\lambda}^2 + v\vec{\lambda}^{2(1-a)} \Big\}
\nonumber \\
&& + \, g^2 w \vec{\lambda}^2 I + \cdots
\, ,
\end{eqnarray}
where
\begin{eqnarray} 
I = \frac{1}{D} \int_{\brm{q}} \sum_{\vec{\kappa}} \frac{\big[1 +  (1-a) h \vec{\kappa}^{-2a} \big]^2 \vec{\kappa}^2}{\big( \tau + \brm{q}^2 + \vec{\kappa}^2 + h \vec{\kappa}^{2(1-a)}\big)^4}
\end{eqnarray}
remains to be evaluated. Upon rewriting $I$ in Schwinger parameterization, the momentum integration can be carried out immediately. The summation over $\vec{\kappa}$ can be approximated by an integration since we have already excluded $\vec{\lambda}= \vec{0}$ properly. Recasting this integration in spherical coordinates we get for $D\to 0$
\begin{eqnarray}
\label{step5}
I &=& \frac{1}{6 \, (4\pi)^{d/2}} \int_0^\infty ds \, s^{3-d/2} \exp \left( -s\tau \right) \int_0^\infty d\kappa \, \kappa
\nonumber \\
&& \times \, \left[ 1 + (1-a) h \kappa^{-2a} \right]^2 \exp \left[ -s\left( \kappa^2 + h \kappa^{2(1-a)} \right) \right] \, .
\nonumber \\
\end{eqnarray}
The integration over radius variable $\kappa$ is simplified by introducing an integration variable $x=s\kappa^2$. Moreover, we apply the binomial formula and expand the rightmost exponential function in Eq.~(\ref{step5}) with the result
\begin{eqnarray}
\label{step6}
I &=& \frac{1}{12 \, (4\pi)^{d/2}} \int_0^\infty ds \, s^{2-d/2} \exp \left( -s\tau \right) 
\nonumber \\
&& \times \, \sum_{k=0}^\infty \sum_{n=0}^2 
\binom{2}{n}
(1-a)^n \, \frac{(-1)^k}{k!} \, h^{n+k} s^{a(n+k)}
\nonumber \\
&& \times \,
\int_0^\infty dx \, \exp (-x) \, x^{k-a(n+k)}\, .
\end{eqnarray}
The integral over $x$ is divergent for $a\geq 1/2$ signifying that the expansion in $\vec{\lambda}^2$ is justified only for $a<1/2$. We assume that $a$ is sufficiently small and abbreviate the value of this integral by $c_{k,n}$. Changing the summation index $k$ to $k-n$ we obtain
\begin{eqnarray}
\label{step7}
I &=& \sum_{k=0}^\infty c_k \, h^k \int_0^\infty ds \, s^{2-d/2+ak} \exp \left( -s\tau \right) \, ,
\end{eqnarray}
where we have set
\begin{eqnarray}
c_k  = \frac{1}{12 \, (4\pi)^{d/2}} \sum_{n=0}^2 
\binom{2}{n} 
(1-a)^n \, \frac{(-1)^{k-n}}{(k-n)!} \, c_{k-n, n} \, .
\nonumber \\
\end{eqnarray}
The integration over $s$ yields
\begin{eqnarray}
\label{step8}
I &=& \sum_{k=0}^\infty c_k \, h^k \, \Gamma \left( \frac{\varepsilon}{2} + ka \right) \tau^{-(\varepsilon /2 + ka)} \, .
\end{eqnarray}
For $a$ of order $\varepsilon$ we can expand the $\Gamma$ function which provides us with the final result
\begin{eqnarray}
\label{step9}
I &=& \sum_{k=0}^\infty \frac{C_k \, h^k}{\varepsilon + 2ka} \, \tau^{-(\varepsilon /2 + ka)} \, ,
\end{eqnarray}
where $C_k$ abbreviates $C_k = 2c_k \Gamma ( 1 + \varepsilon /2 + ka )$. We see that
the UV singularities manifest themselves in a series of poles of the from $h^k/(L \varepsilon + 2ka)$ with $L=1$. In other words, our one-loop example is in agreement with the general insights presented in Sec.~\ref{RGA_RSC}. 



\begin{thebibliography}{}
\bibitem{footnote1}
By virtue of Eq.~(\ref{fDef}), the average resistance of an occupied bond is given by $(1-a)\sigma_0^{-1}\int_0^1 dx \, x^{-1-a}$ and hence diverges for $a \geq 0$.
\end{thebibliography}
\end{document}